 \documentclass[12pt]{iopart}

\usepackage{hyperref,graphicx}

\begin{document}

\title{\bf Binary continuous random networks}

\author{Normand Mousseau\dag\ and G. T. Barkema\ddag}

\address{\dag\ D\'epartement de physique and Regroupement qu\'eb\'ecois sur
  les mat\'eriaux de pointe, Universit\'e de Montr\'eal, C.P. 6128,
  succ. centre-ville, Montr\'eal (Qu\'ebec) Canada H3C 3J7}

\address{\ddag\ Theoretical Physics, Utrecht University,
Leuvenlaan 4, 3584 CE Utrecht, The Netherlands}

\eads{\mailto{normand.mousseau@umontreal.ca}, \mailto{G.T.Barkema@phys.uu.nl}}

\begin{abstract}
  Many properties of disordered materials can be understood by looking at
  idealized structural models, in which the strain is as small as is
  possible in the absence of long-range order. For covalent amorphous
  semiconductors and glasses, such an idealized structural model, the
  continuous-random network, was introduced 70 years ago by
  Zachariasen. In this model, each atom is placed in a crystal-like
  local environment, with perfect coordination and chemical ordering,
  yet longer-range order is nonexistent. Defects, such as missing or
  added bonds, or chemical mismatches, however, are not accounted for.
  In this paper we explore under which conditions the idealized CRN
  model without defects captures the properties of the material, and
  under which conditions defects are an inherent part of the idealized
  model. We find that the density of defects in tetrahedral networks
  does not vary smoothly with variations in the interaction strengths,
  but jumps from close-to-zero to a finite density.  Consequently, in
  certain materials, defects do not play a role except for being
  thermodynamical excitations, whereas in others they are a
  fundamental ingredient of the ideal structure.
\end{abstract}

\maketitle

\section{Introduction}

Since the work of Zachariasen~\cite{zachariasen32}, the continuous random
network (CRN) model has been considered an idealized yet reasonable
representation of oriented glasses and amorphous semiconductors
such as silica and chalcogenide glasses, amorphous silicon and
amorphous gallium arsenide. As such, these idealized networks
were the subject of extensive studies during the 1970's and the
1980's~\cite{bell66,henderson74,guttman80}, including by a number of
researchers present in this Workshop celebrating the 60th anniversary
of Mike Thorpe. It turns out that, even though it is easy to assemble
a generic CRN, it is much more difficult to create a network that
minimizes the strain and yet satisfies fully the basic constraints
of coordination. Much effort, therefore, has gone in simply generating
high-quality ---i.e., low strain--- CRNs and comparing their properties
to well-annealed {\it a}-Si or {\it g}-SiO$_2$.

Let us consider the simplest CRN model, a tetrahedral network
consisting of elements connected in such a way that each element has
exactly four bonds. This network represents the ideal state for
elemental amorphous semiconductors such as silicon, germanium and
binary alloys such as SiGe, GaAs and InP. Experimentally, this state is not
uniquely defined and depends strongly on the method of preparation as
well as on the annealing history. However, well-annealed samples are
thought to be represented by low-strain CRNs.

This relation is not universally accepted as the model suffers from
some limitations. In particular, the CRN model does not allow defects
which could play an important role in decreasing strain in amorphous
networks.  High-Q measurements of the radial distribution function in
{\it a}-Si generated by ion-implantation~\cite{Laaziri} indicate, for
example, an average coordination of 3.88, significantly below 4.0. The
question of defects is possibly even more important when considering
binary semiconductors, such as {\it a}-GaAs, where the strain energy
is in competition with the chemical energy and for which two types of
defects can then be generated: coordination and chemical defects.

This additional competition makes the generation of low-strain
binary-alloy models even more difficult and relatively little work has
been done on these systems. Connell and Temkin~\cite{connell74}
produced a 64-atom CRN with periodic boundary conditions and no
odd-membered rings, demonstrating that a chemically-ordered CRN was
possible.  Recently, Mousseau and Lewis~\cite{mousseau97a,mousseau97b}
generated models with a low density of coordination and chemical
defects, showing that the cost of a chemically unfavorable bond in
GaAs was sufficient to strongly favor even-membered rings.

The models generated in the studies by Mousseau and Lewis had some
coordination and chemical defects, however, leaving open the question
as to whether it is possible to construct a CRN with only
even-membered rings, and still the same strain level as a
non-constrained CRN. In this Paper, we revisit this question using a
modified version of the celebrated bond-switching algorithm of Wooten,
Winer and Weaire~\cite{www,ww} to generate CRNs with a specific
density of coordination and chemical defects.

To study the role of defects in disordered materials, we start with
the ordinary tetrahedral CRN model, a high-quality and well-defined
reference state. Two types of defects are then introduced: {\it dangling
bonds}, associated with undercoordinated atoms, and {\it wrong bonds},
defined in this case as a bond between two atoms of the same species in
a binary alloy. Dangling bonds are found in any elemental and alloyed
amorphous semiconductors; wrong bonds belong only to the later type,
GaAs or InP, for example.

Using a simple quadratic potential,~\cite{keating66} we find a linear
relation between the density of defects and the strain in the lattice
for the lowest-energy structures. This suggests that chemical defects
are either very rare or exist with a density of at least 16 \% since the energy gain of adding or removing a defect is the same at all concentrations.



\section{Computer generation of well-relaxed CRN}

Because of its link with a simple elemental material, the most studied
CRN model has been the tetrahedral one.

The first method for producing efficiently, and in a reproducible
fashion, numerical models of CRN was introduced in 1985.  The 1970s
had seen a number of hand-built models with a large surface-to-volume
ratio and, in 1980, Guttman proposed a first numerical algorithm for
preparing tetrahedral CRNs~\cite{guttman80}. However, this algorithm
only worked for small cells of less than 100 atoms and was not easily
implemented. Assembling three ``W'''s a few years before their time,
Wooten, Winer and Weaire~\cite{www,ww} devised a clever algorithm,
involving local bond switches, that could rapidly transform a
crystalline diamond structure into a low-strain amorphous structure.

This method, and the original models, were used by many groups over the
following decade to study flexibility and other structural and electronic
properties of fully-coordinated {\it a}-Si networks as well as related
materials such as SiO$_2$ and {\it a}-Si:H.

Twenty years after its introduction, the WWW algorithm remains a useful
tool. Improving computational aspects of the algorithm, we could gain
about two orders of magnitude in efficiency~\cite{barkema00,vink01},
allowing us to generate low-strain models of {\it a}-Si with up to 100,000
atoms and {\it v}-SiO$_2$ up to 300,000 atoms~\cite{vink03}, as well as
study paracrystalline Si~\cite{nakhmanson01}, a controversial new phase
identified recently.

\subsection{The WWW bond-switching algorithm}

\begin{figure}
\begin{center}
\includegraphics[width=8cm]{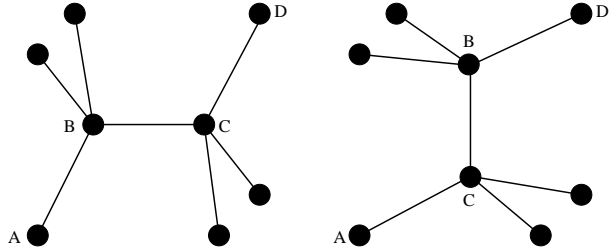}
\end{center}
\caption{\label{fig:wwwmove} Diagram depicting the WWW bond transposition.
Four atoms A, B, C, and D are selected following the geometry shown left;
two bonds, AB and CD, are then broken and atoms A and D are reassigned
to C and B, respectively, creating two new bonds, AC and BD, resulting
in the geometry shown right.}
\end{figure}

In the WWW approach, a configuration consists of the coordinates of
$N$ atoms and a list of the $2N$ bonds between them. The structural
evolution consists of a sequence of bond transpositions as illustrated
in Fig.~\ref{fig:wwwmove}: in a chain of four bonded atoms, ABCD, the
two bonds AB and CD are replaced by two other bonds AC and BD, leading
to a chain ACBD after the bond transposition.  The generation of a CRN
starts with a cubic diamond structure which is randomized by a large
number of such bond transpositions. After randomization, the network
is relaxed through a sequence of bond transpositions, accepted with
the Metropolis acceptance probability~\cite{metrop}:
\begin{equation}
\label{eq:metropolis}
  P = \min \left[1, \exp \left( \frac{E_b-E_f}{k_{\mathrm{B}} T}
    \right)  \right],
\end{equation}
where $k_{\mathrm{B}}$ is the Boltzmann constant, $T$ is the temperature,
and $E_b$ and $E_f$ are the total quenched energies of the system before
and after the proposed bond transposition.

With an explicit list of neighbors, it is possible to use a simple harmonic
interaction such as the Keating potential~\cite{keating66} to calculate
energy and forces:
\begin{eqnarray}
\label{eq:keating}
  E &=& \frac{3}{16} \frac{\alpha}{d^2} \sum_{<ij>}
        \left( \vec{r}_{ij} \cdot \vec{r}_{ij} - d^2\right)^2 \nonumber\\
    &+& \frac{3}{8} \frac{\beta}{d^2} \sum_{<jik>}
        \left( \vec{r}_{ij} \cdot \vec{r}_{ik} + \frac{1}{3} d^2 \right)^2,
\end{eqnarray}
where $\alpha$ and $\beta$ are the bond-stretching and bond-bending force
constants, and $d=2.35$ \AA\ is the Si-Si strain-free equilibrium bond
length in the diamond structure. For Si, the standard values for the force
constants are $\alpha=2.965$ eV \AA$^{-2}$ and $\beta=0.285\alpha$.
Since the list of interacting atoms is explicit and not based on the
interatomic distance, it is possible for two atoms to be virtually on top of
each other but still not interact if they are not explicitly bonded.
This unwanted situation becomes however increasingly rare with improved
quality of the CRN structure.

Wooten and Weaire followed this approach to generate a 216-atom model
with an angular distribution of 10.9 degrees~\cite{ww}. A decade later,
Djordjevi\'c, Thorpe, and Wooten exploited the advance in computing
hardware and produced two large 4096-atom networks with a bond-angle
distribution of 11.02 degrees for configurations without four-membered
rings and 10.51 degrees when these rings are allowed~\cite{djordjevic95}.

The WWW algorithm in its original form is not well suited to generate
CRNs much larger than a few thousand atoms. This is mostly due to the
fact that for each proposed bond transposition, about one hundred energy
and force calculations are required, each scaling as $\cal O$$(N)$ with
system size $N$. These $\cal O$$(N)$ operations are the bottleneck of
the algorithm.  A few years ago, we presented a number of modifications
to the original WWW algorithm, partially aimed at resolving these poor
scaling properties~\cite{barkema00}.

Using the improved WWW algorithm, we generated two 1000-atom models
with bond angle deviations as low as 9.20 degrees~\cite{barkema00}.
Furthermore, using the same algorithm we generated a 4096-atom model
with an angular deviation as low as experimentally accessible -- around
nine degrees. With some more improvements in the implementation of the
algorithm, we succeeded in generating well-relaxed configurations of
up to 100,000 atoms~\cite{vink01,vink03}.

\section*{Extension to binary networks with defects}

The original WWW algorithm was designed to generate generic tetrahedral
CRNs.  Here, we introduce a series of modifications to this algorithm that
allow us to control the degree of chemical ordering and the proportion
of coordination defects.

The crystalline structure of binary networks does not provide a good
starting point, since the temperatures at which it starts to melt in a
reasonable time are so high that, before the crystalline structure has
been left behind everywhere, some parts of the network develop lots of
unwanted anomalies which require a lot of effort to anneal. We therefore
construct initial CRNs with a different procedure, guaranteed to have no
remainders of the crystalline state.  Atoms are initially placed in the
box at random, and labeled A and B, in equal proportions. Bonds are then
assigned to pairs of differently labeled atoms, with a strong preference
for near atoms, until the desired total coordination of four is reached.
The elastic strain, defined by the Keating potential (Eq.~\ref{eq:keating})
plus a repulsive constant in the case of wrong bonds, is then minimized
under fixed topology.  This whole procedure results in initial CRN
configurations with a bond-angular spread of around 35 degrees, without
long-range order, nor chemical or coordination defects.

\begin{figure}
\begin{center}
\includegraphics[width=8cm]{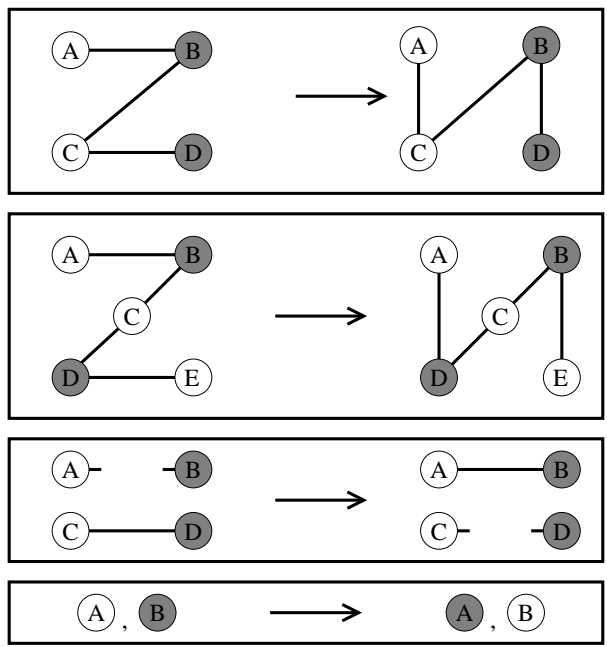}
\end{center}
\caption{\label{fig:gaasmove} Diagrammatic representation of the moves
used in our simulations on binary CRNs. a) We employ the usual bond
transposition as proposed by Wooten, Winer and Weaire.  This move
introduces (and sometimes removes) chemical defects, as can be seen
by the appearance of the bonds between two white atoms and between
two grey atoms. b) If instead of two nearest-neighbor atoms, two
next-nearest-neighbor atoms exchange their neighbors, a sample without
chemical defects stays chemically perfectly ordered. c) Simultaneously a
pair of dangling bonds is bonded, while a bond is removed elsewhere. d)
Two chemically different atoms exchange their chemical nature.}
\end{figure}

The topology of these initial configurations is then relaxed through
a series of local and non-local Monte Carlo moves, as described in
Figure~\ref{fig:gaasmove}.  Some of these moves can introduce chemical
defects, if the move results in a sufficient reduction in the elastic
strain; although the initial configuration is chemically ordered, the
chemical state of the final configuration depends on the cost assigned
to wrong bonds.  A thorough relaxation for 1000-atom models requires
several million attempts per atom, which can be done in a few days on
a fast workstation.

On a tetrahedral CRN without penalty for odd-membered rings, a smart
assignment of the atomic labels results in a defect density around
15\%; a lower density of defects can only be obtained by chemical
ordering leading to a lower proportion of odd-membered rings in the
network. 

\section{Results}

We generated well-relaxed configurations with a varying degree of
chemical ordering by tuning the energetic penalty for chemical
defects.  All the configurations used in this work are 1000-atom cells
at the density of the diamond crystal structure, i.e., near zero
pressure. Depending on the ratio of the potential parameters tuning
the bond length and angles, there is a limited trade-off between these
two quantities.  To be physically relevant, the bond-length
fluctuation in these models should not exceed a few percent. In this
case, the width of the bond-angle distribution becomes a good
indicator for structural properties of the networks, scaling linearly
with the total energy, irrespective of the details of the interaction
potential. 

Without any defect, the best model we can construct still shows a
bond-angle distribution of 15 degrees, much higher than the 9 to 10
degrees that can be reached in a model without chemical order.

\begin{figure}
\begin{center}
\begin{tabular}{c}
\includegraphics[width=8cm]{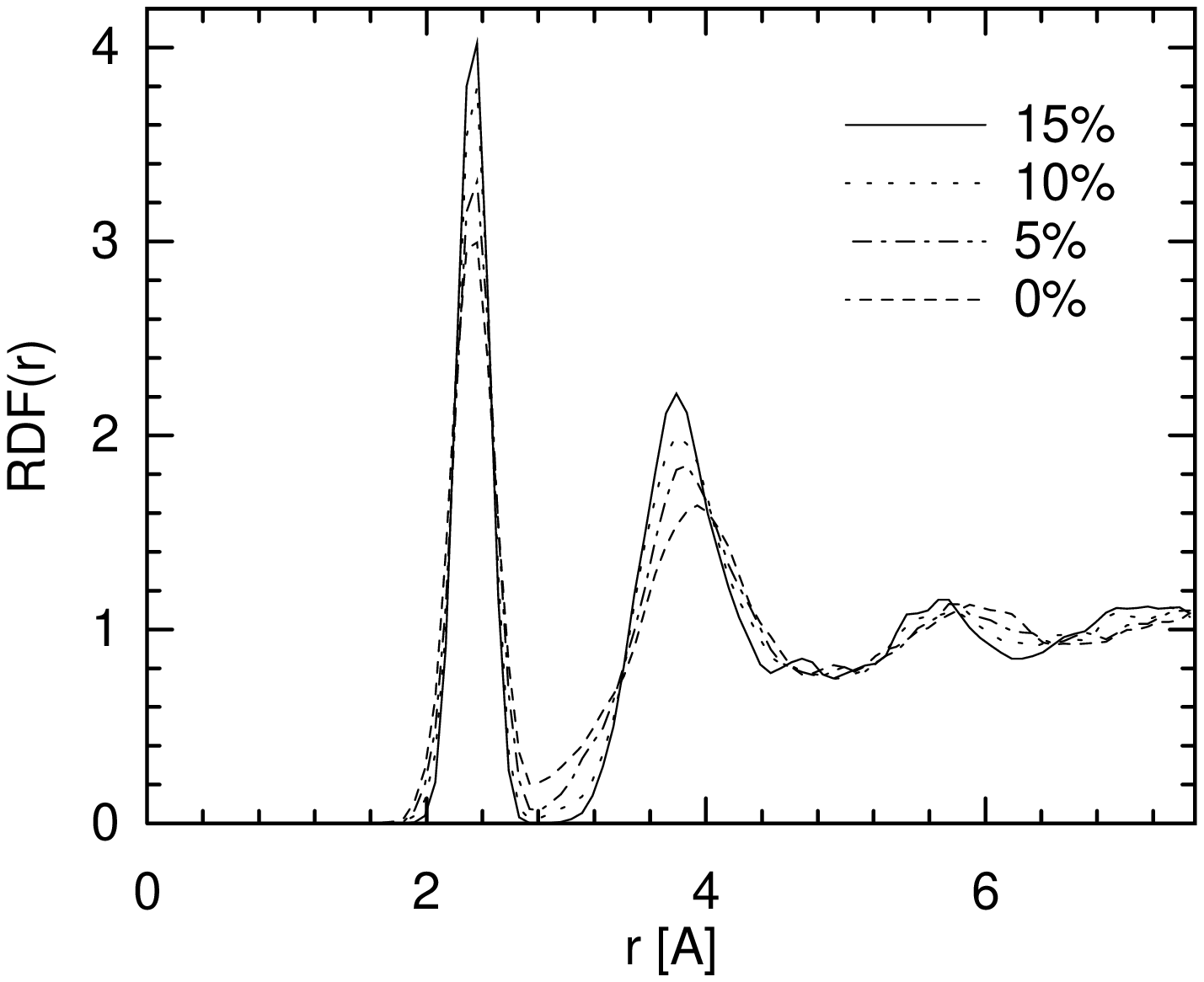}\\
\includegraphics[width=8cm]{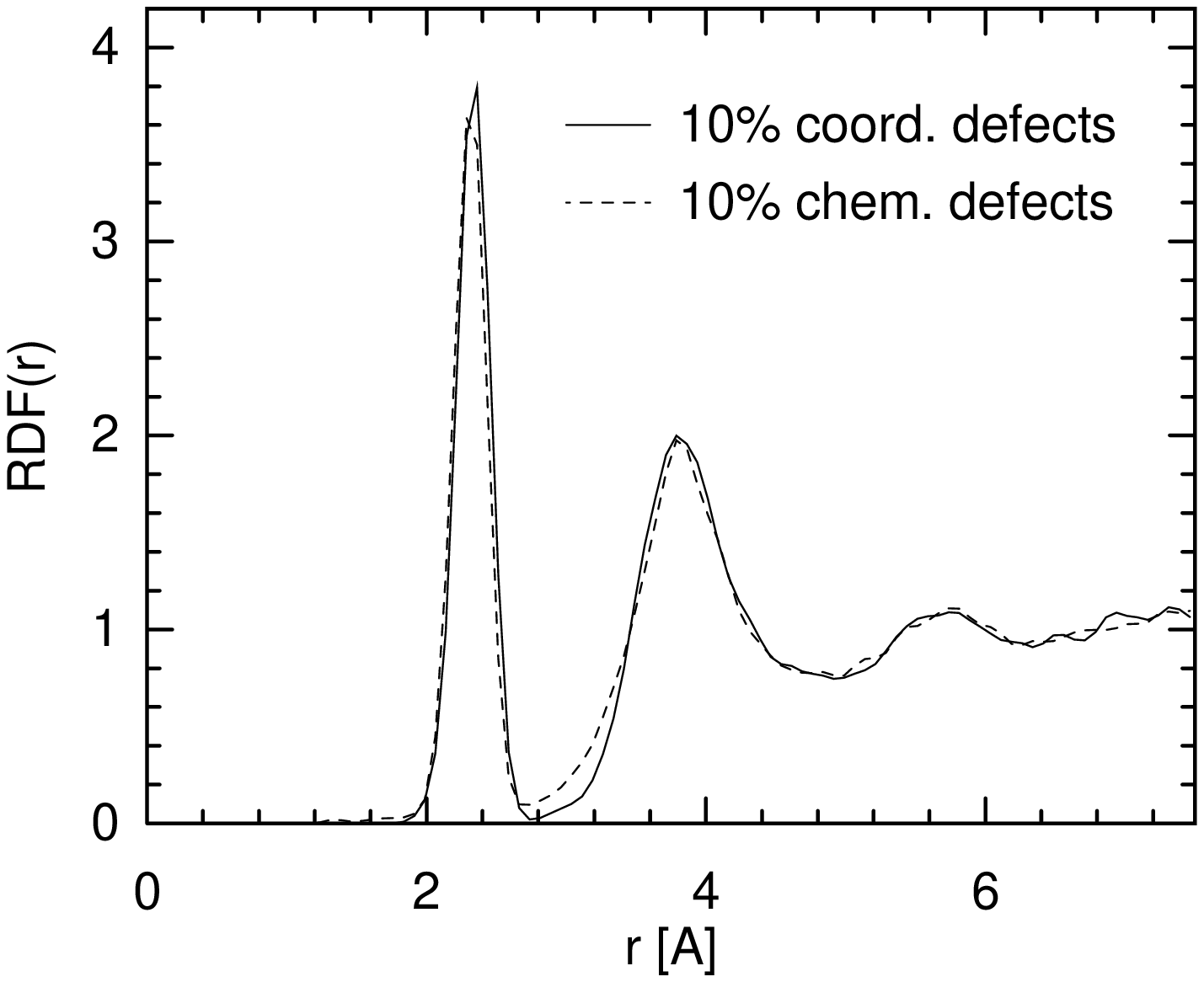}
\end{tabular}
\end{center}
\caption{ 
Top: Radial distribution functions obtained from the samples without
coordination defects, but with 0\%, 5\%, 10\% and 15\% chemical defects
(wrong bonds).  Bottom: Comparison of radial distribution functions
obtained with samples with either 10\% chemical defects, or 10\%
coordination defects.
}
\label{fig:rdf}
\end{figure}

As mentioned above, with about 15\% chemical defects and suitable
labeling, the network topology is equivalent to that of {\it a}-Si.
Similary, 15\% of under-coordination is sufficient to ensure no chemical
defect with ring statistics identical to that of {\it a}-Si with the same
density of defects. The similarity between these two defects can be seen
in Fig.~\ref{fig:rdf}, which shows the radial distribution function (RDF)
for a perfectly ordered and coordinated lattice as well as for lattices
with 5, 10 and 15\% chemical defects. The bottom panel compares the RDFs
of two models with 10 \% of chemical and coordination defects. Even
though the RDF changes as a function of defects, as was demonstrated
previously~\cite{mousseau97a}, this is not a very sensitive quantity and
it is not precise enough to provide a well-defined experimentally-measured
defect density.

\begin{figure}
\begin{center}
\begin{tabular}{c}
\includegraphics[width=8cm]{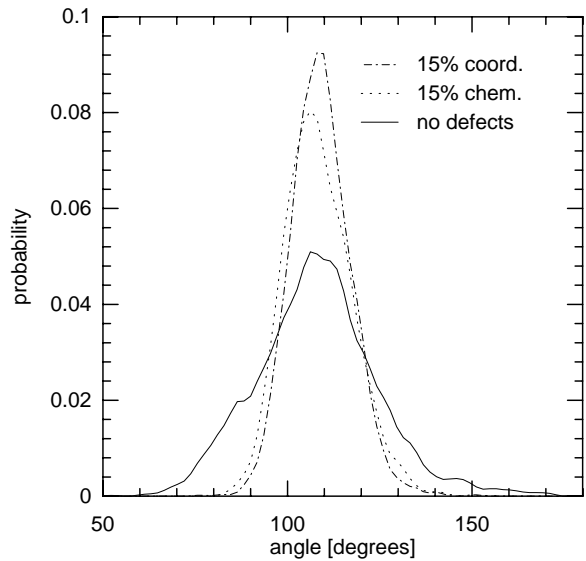}
\end{tabular}
\end{center}
\caption{Bond angle distribution for the perfectly coordinated and
chemically ordered model with a width of $\Delta \theta=16.42^\circ$,
with 15\% chemical defects ($\Delta \theta= 9.16^\circ$), and with 15\%
coordination defects ($\Delta \theta =8.12^\circ$).
}
\label{fig:angle}
\end{figure}

This is confirmed by plotting the bond-angle distribution as shown in
Fig.~\ref{fig:angle}.  We could have expected that coordination defects
would be better at removing the strain in the model, as each defect also
removes a constraint on the bond-length, yet the effect is surprisingly
small.  The two systems relax with a bond-angle distribution width of
8.12 and 9.16 degrees.

\begin{figure}
\begin{center}
\vspace*{-1.6cm}
\includegraphics[width=12cm]{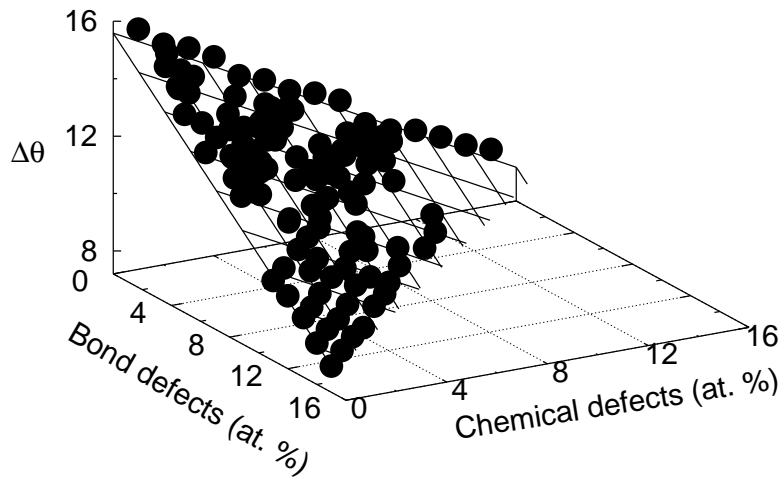}
\end{center}
\caption{Minimal angular spread obtained in our binary tetrahedral
networks, as a function of the density of coordination and chemical
defects.}
\label{fig:phasediagram}
\end{figure}

The main result is presented in Figure \ref{fig:phasediagram}, where the
minimal angular spread is plotted as a function of the density of
coordination and chemical defects. The resulting surface defines the
``optimal'' surface, i.e. a characterization of networks with the lowest
elastic strain. Networks above this optimal surface will eventually
evolve to this surface by reducing the number of coordination or chemical
defects, or by reducing the elastic strain. It should not be possible
to find experimental samples falling well below this surface.

Excepting densities below a percent, where our relaxation algorithm is
less efficient, there is a linear relation between these two
quantities, and thus between this defect density and the total energy;
the surface is well fitted by a plane, described by
\begin{equation}
\Delta \theta = a -b \rho_c -c \rho_t,
\end{equation}
where $\rho_c$ is the density of chemical defects, and $\rho_t$ is the
density of topological defects. A least-squares fit to the parameters
$a$, $b$, and $c$ gives $a=15.59 \pm 0.05$, $b=49.8 \pm 0.5$ and $c=45.1
\pm 0.7$.

Apparently, the release in elastic strain per defect added is
constant, i.e., it is equally hard to remove the first defect as the
last one. If the nature of the interaction strengths in the material
is such that it can remove a single defect, it will be strong enough
to remove {\it all} defects!

\section{Conclusion}

Generating a set of well-relaxed configurations with various densities
of chemical and coordination defects, we studied the impact of
chemical ordering and bonding on the structural properties of
continuous random networks. For this, we use an enhanced WWW-algorithm
with millions of attempted moves per atom. In view of this extensive
sampling, the configurations we obtain should be close to the optimal
ones in terms of strain, allowing the following conclusions.

First, a perfectly coordinated and chemically ordered binary CRN
cannot be relaxed to a structure with a bond-angle distribution below
about 15 degrees. Comparing with other amorphous networks, this is a
high degree of strain. It is unlikely, therefore, that a materials
such as {\it a}-GaAs can be found without a sizable density of
defects. 

Interestingly, if the sample cannot tolerate this kind of strain, then
it should display about 15 \% of defects--- either coordination or
chemical--- as the relation between strain and defects is linear.
Should a single defect be preferable to none, then the sample cannot
be stable at any defect density below 15\%. At this defect density,
the strain becomes equal to or slightly below that of an elemental
CRN.

These results hinge of course crucially on the issue whether the CRN
model is a reasonable description of amorphous binary semiconductors.
For elemental amorphous semiconductors, it is possible to generate
perfect CRN models with a strain density equal to annealed samples. No
such information is available for binary semiconductors.  Experimental
measurements to establish whether the CRN model can be used for studying
the structure of these materials would be of great interest, as direct
structural values offer little precision on defect densities in these
systems.

\ack

NM is supported in part by the {\it Fonds qu\'eb\'ecois de la recherche
  sur la nature et les technologies}, the {\it Natural Sciences and
  Engineering Research Council} of Canada and the Canada Research
Chair Program. NM is a Cottrell Scholar of the Research Corporation.


\section*{References}
\begin{thebibliography}{99}

\bibitem{zachariasen32}
Zachariasen WH, J. Am. Chem. Soc.
1932;54:3841.

\bibitem{bell66} Bell RJ and Dean  P, Nature (London)
1966; 212:1354.

\bibitem{henderson74} Henderson D,  J. non-Cryst. Sol. 
1974; 16:317.

\bibitem{guttman80} Guttman, L., Ching, W. Y., and Rath, J., 1980, Phys.
Rev. Lett. {\bf 44}, 1513.

\bibitem{Laaziri}
Laaziri K, Kycia S, Roorda S, Chicoine M, Robertson JL, 
Wang J and Moss SC, Phys. Rev. Lett. 1999; 82:3460.

\bibitem{connell74} G. A. N. Connell and R. J. Temkin,
Phys. Rev. B {\bf 9}, 5323 (1974).

\bibitem{mousseau97a} Mousseau, N. and Lewis, L.J., 1997, Phys. Rev. Lett.
{\bf 78}, 1484.


\bibitem{mousseau97b} Mousseau, N. and Lewis, L.J., 1997, Phys. Rev. B
  {\bf 56}, 9461

\bibitem{www}
Wooten F, Winer K and Weaire D, Phys. Rev. Lett. 1985; 54:1392.

\bibitem{ww} Wooten F and Weaire D,  Solid State Phys.  1987; 40:1.

\bibitem{barkema00} Barkema, GT and Mousseau N, Phys. Rev. B.
2000; 62:4985.

\bibitem{vink01}  Vink RLC, Barkema GT,  Stijnman M and Bisseling RH,
Phys. Rev. B 2001; 64:245214.

\bibitem{vink03}  Vink RLC and Barkema GT, Phys. Rev. B 2003; 67:245201.






\bibitem{metrop} Metropolis N, Rosenbluth AW, Rosenbluth MN, Teller AH, and
Teller E, J. Chem. Phys. 1953; 21:1087.

\bibitem{keating66} Keating PN, Phys. Rev. 1966; 145:617.

\bibitem{djordjevic95}
Djordjevi\'c BR, Thorpe MF and Wooten F,  Phys. Rev. B
1985; 52:5685.






\bibitem{nakhmanson01} Nakhmanson SM, Voyles PM, Mousseau N,
Bar\-ke\-ma GT, and Drabold DA, Int. J. Mod. Phys. B 2001; 15:3253;
Phys. Rev. B 2001; 63:235207. 

























\bibitem{models} The models discussed in this Review are available freely
from either author.


\end{thebibliography}
\end{document}